\documentclass{aastex63}
\usepackage{amsmath,ulem,mleftright}
\usepackage{mathtools}
\usepackage{flexisym}
\usepackage{comment}
\usepackage{xparse}
\NewDocumentCommand{\evalat}{sO{\big}mm}{%
  \IfBooleanTF{#1}
   {\mleft. #3 \mright|_{#4}}
   {#3#2|_{#4}}%
}
\usepackage[toc,page]{appendix}
\usepackage[caption = false]{subfig}
\usepackage{natbib}
\begin{document}
\title{Investigating toroidal flows in the Sun using normal-mode coupling}
\author{Prasad Mani \& Shravan Hanasoge}
\affiliation{Department of Astronomy $\&$ Astrophysics, Tata Institute of Fundamental Research, Mumbai 400005, India}
\email{prasadmani94@gmail.com}
\begin{abstract}
Helioseismic observations have provided valuable datasets with which to pursue the detailed investigation of solar interior dynamics. Among various methods to analyse these data, normal-mode coupling has proven to be a powerful tool, used to study Rossby waves, differential rotation, meridional circulation and non-axisymmetric multi-scale subsurface flows. Here, we invert mode-coupling measurements from Helioseismic Magnetic Imager (HMI) and Michelson Doppler Imager (MDI) to obtain mass-conserving toroidal convective flow as a function of depth, spatial wavenumber and temporal frequency. To ensure that the estimates of velocity magnitudes are proper, we also evaluate correlated realization noise, caused by the limited visibility of the Sun. We benchmark the near-surface inversions against results from Local Correlation Tracking (LCT). Convective power likely assumes greater latitudinal isotropy with decrease in spatial scale of the flow. We note an absence of a peak in toroidal-flow power at supergranular scales, in line with observations that show that supergranulation is dominantly poloidal in nature.

\end{abstract}

\section{Introduction} \label{sec:intro}
Helioseismology has been successful at imaging the solar interior by studying oscillations at the surface \citep[][]{CD2002}. It has been exercised to learn about a wide variety of features in the Sun such as magnetism \citep[][]{Duvall2000, Jensen2001, Birch2010, Birch2013}, convection \citep[][]{Miesch, SandN2009, Han2012, Han2016}, differential rotation \citep[][]{Schou1998}, meridional circulation \citep[][]{Giles1997} and Rossby waves \citep[][]{Loptein2018, Liang2019}.
Several techniques such as global-mode helioseismology \citep[e.g.,][]{CD2002}, ring-diagram analysis \citep[][]{Hill1988}, time-distance helioseismology \citep[][]{Duvall1993}, helioseismic holography \citep[see][for a review]{Lindsey2000}, direct modelling \citep[][]{Wood2002}  have proven to be successful in imaging the solar interior. Normal-mode coupling, \citep[][]{Wood1989, LandR1992} the method of choice here, uses near-resonance correlations of the surface measurements of solar oscillations in the wavenumber-frequency domain,  permitting inferences of time variations of non-axisymmetric features.

The first use of mode-coupling in helioseismology goes back to \citet{Wood1989}, and subsequently by \citet{LandR1992}, who adapted the mature geophysical development of the technique to helioseismic problems, and more recently by \citet[][]{Roth2008, Vor2007, Vor2011, Wood2007, Wood2014, Wood2016, Roth2011, Roth2020}. Utilizing the algebra of mode coupling detailed in \citet{Han2017} and \citet{Han2018}, \citet{HandM2019} and \citet{MandH2020} detected and studied properties of Rossby waves. An important test of mode coupling was the recent analysis of \citet{Samarth2021} to infer solar differential rotation and \citet{Hanson2021} to analyze the power spectrum of supergranules, two well established results, thereby validating the method.

As with all other helioseismic techniques, mode coupling proceeds by assuming the structure and dynamics present in the Sun to be modelled as small perturbations to the reference model, which suffers from the absence of rotation, flows and magnetism \citep[we use Model S][]{CD1996}. We can express the solar oscillation eigenfunctions as a weighted linear sum of model eigenfunctions. The associated weights, known as coupling coefficients, encode properties of the solar interior. We then state the linear forward problem derived using mode coupling by relating the coupling coefficients and the observed wavefield correlations to the underlying perturbations.

\citet{Han2017} reevaluated and extended the results of \citet{LandR1992} and \citet{Han2018} accounted for systematical errors arising from the partial visibility of the Sun (spectral leakage) in mode-coupling measurements, attempting to model the observations better and improve the accuracy of flow velocity inferences in \citet{HanSciA2020}. They found that toroidal flows on large scales are confined to the equatorial regions. To further validate mode coupling as a tool with which to probe convection, \citet{Mani2020} carried out extensive tests to validate inversions of synthetic mode-coupling measurements for toroidal flow while accounting for leakage, and concluded that inversions are strongly influenced by the model assumed for the correlation between flow velocities. 

In this work, we extend the analysis of spatial scales of the toroidal flow and characterize velocities and power speectra. To this end, we examine 4 yr Helioseismic Magnetic Imager \citep[HMI;][]{HMI2012} and Michelson Doppler Imager \citep[MDI;][]{MDI1995} data each and perform inversions using Regularized-Least-Squares (RLS) and Subtractive Optimally Localized Averages (SOLA) \citep[][]{PandT1994}. We consider leakage in measurements and also evaluate correlated realization noise using the model derived in \citet{Han2018}. As an independent yardstick for the velocities obtained from mode-coupling, we compare inversions at the surface ($\sim0.995R_{\odot}$) with velocities obtained from Local Correlation Tracking \citep[][]{Loptein2018}. LCT obtains horizontal velocities by maximizing the correlation of granulation pattern in successive intensity images - the idea being that the pattern is advected by flows.

For a vector flow field ${\bf u}_{o}^{\sigma}({\bf r})$, where \textbf{r} denotes the 3-dimensional spatial co-ordinate and $\sigma$ the flow evolution scale, the Chandrasekhar-Kendall decomposition gives the toroidal (vortical) part of the flow as 
\begin{equation}\label{flow}
    {\bf u}_{o}^{\sigma}({\bf r}) = \sum_{s,t} \;w_{st}^{\sigma}(r)\hat{\bf{r}}\times{\boldsymbol\nabla}Y_{s}^t,
\end{equation}
where $w_{st}^{\sigma}(r)$ are the toroidal-flow coefficients which we seek to infer from measurements, $Y_{st}$ is the spherical harmonic of degree $s$ and the azimuthal order $t$ and $r$ is radius. Toroidal flows are horizontal flows with curl but no divergence, and hence by definition, mass conserving. We ignore $t=0$ as we are only interested in non-axisymmetric flows.

\section{Data Analysis}
We consider global-mode time-series with $\ell\in\big[50,180\big]$ and the resonant frequency of the mode $\nu_{n\ell}\in\big[1.8,3.6\big]$mHz, where the indices ($n,\ell,m$) denote radial order, spherical-harmonic degree, and the azimuthal order of a mode, respectively. Modes of higher frequencies are known to show systematical errors \citep[][]{Antia2008}, and on the lower frequency end, sampling the time series data off-resonance causes background power to leak in due to the small line widths of these modes. Denoting the temporal Fourier transform of the series by $\phi_{\ell m}^{\omega}$ where $\omega$ is temporal frequency, we compute mode-coupling measurements as $\phi_{\ell m+t}^{\omega+\sigma}\phi_{\ell m}^{\omega*}$. Correlating wavefields at same-spherical-harmonic degree, different azimuthal orders and temporal frequencies renders the measurement sensitive only to odd-degree (odd-$s$) toroidal flow \citep[][]{Han2018, Mani2020}. To facilitate data analysis, we condense $\phi_{\ell m+t}^{\omega+\sigma}\phi_{\ell m}^{\omega*}$ into its linear-least-square fit \citep[][]{Wood2016}, known as $B$-coefficients, given by
\begin{equation}\label{leastsquaresfit}
    B_{st}^{\sigma}(n, \ell) = \sum\limits_{m, \omega}W_{\ell m s t}^{\omega+\sigma}\phi_{\ell m+t}^{\omega+\sigma + t\Omega}\phi_{\ell m}^{\omega*}
\end{equation}
where $W$ is theoretically derived in the \citet{HanSciA2020} supplementary and also in \citet{Han2018}. These measurements are taken in a rotating frame, at the equatorial rate $\Omega=453$nHz. This leads to the transformation $\sigma\rightarrow\sigma + t\Omega$, which is incorporated in equation~(\ref{leastsquaresfit}).

The forward model linearly connects perturbations in internal solar properties with respect to a solar model to observations. In the first-Born approximation, the wavefield correlation is linearly related to the flow through a sensitivity kernel \citep[][]{Wood2006}. Using the forward model derived through mode-coupling in \citet{Han2018}, the inverse problem is stated as
\begin{equation}\label{Bst}
    B_{st}^{\sigma}(n,\ell) = \sum_{s't'}\int_{\odot} dr\: w_{s't'}^{\sigma}(r)\Theta_{st}^{s't'}(r;n,\ell,\sigma).
\end{equation}
The sum over $s',t'$ indicates that leakage between oscillation signals translates to leakage between neighboring flow wavenumbers. Hence, to accurately estimate the signal in a given channel ($s,t$), contributions from neighboring modes, weighted by leakage matrices \citep[see][for how leakage is modeled]{SandB1994} in the sensitivity kernel $\Theta_{st}^{s't'}(r;n,\ell)$, need to also be taken into account. {\citet{MandH2020} studied properties of odd-$s$ Rossby modes in the frequency-bin range $\sigma\in[0,2]\mu$Hz and concluded that the dominant leakage into a desired mode is from the neighboring odd-$s$; in the absence of tracking, leakage occurs at the same temporal frequency ($\sigma_{s}\rightarrow\sigma_{s}$), and when tracking is applied, it leaks into higher temporal frequencies ($\sigma_{s}\rightarrow\sigma_{s}+2\Omega$, where $\Omega=453$nHz). Hence we restrict the analysis to $[0,0.9]\mu$Hz.}

Turbulent convection stochastically excites waves with random phase and amplitudes all over the solar surface. If the Sun had no perturbations (i.e., spherical harmonics were the correct eigenfunctions) and if we could observe the entire solar globe, we would be able to perfectly resolve the excited modes into unique spherical harmonic wavenumbers and model them as uncorrelated \citep[see][supplementary]{Han2018,HanSciA2020}. But spatial windowing results in convolutions in spectral space, inducing leakage in the observed modes, which in turn causes them to be finitely correlated. Here, we simply produce the final noise model (i.e., $\langle|B_{st}^{\sigma}|^2\rangle$ for $B$-coefficients obtained from pure noise) 
\begin{equation}\label{noise}
 \epsilon_{st}^{\sigma}(n, \ell) = \sum\limits_{m, \omega}|W_{\ell m s t}^{\omega + \sigma}|^2\langle|\phi_{\ell m+t}^{\omega+\sigma}|^2\rangle\langle|\phi_{\ell m}^{\omega}|^2\rangle.
\end{equation}
To summarise, for both the instruments, 
\begin{enumerate}
    \item using Equation~(\ref{leastsquaresfit}) and Equation~(\ref{noise}), we compute $B$-coefficients and noise \big(for $(n,\ell)$ such that $\ell\in\big[50,180\big]$ and $\nu_{n\ell}\in\big[1.8,3.6\big]$mHz, and $\sigma\in[0,0.9]\mu$Hz\big)
    \item we compute the sensitivity kernels $\Theta_{st}^{s't'}(r;n,\ell)$ only at $\sigma=1\mu$Hz \citep[see Figure 1 of][]{Mani2020}
    \item invert Equation~(\ref{Bst}) for $w_{st}^{\sigma}(r)$ \big(\textit{for all odd $s\in[1,149]$ and $t\neq0$ for desired depth}\big)
\end{enumerate}

\section{RLS inversions}\label{RLS inversions}
Linear inversion methods provide estimates of toroidal flow velocity $w_{st}^{\sigma}$ at the desired depth $r_0$, using a linear combination of observations $B_{st}^{\sigma}$. We employ two methods for proper comparisons - we describe the Regularized-least-squares (RLS) technique here. For SOLA, see Appendix \ref{SOLA Inversions}.

RLS aims at obtaining solutions to inverse problems by navigating the trade-off between the magnitude of the regularized solution and quality of fit to the observed data. The conventional way to achieve optimality is by plotting the goodness of fit against the solution magnitude for different regularization values, resulting in an ``L curve" (Figure~\ref{fig_avgk}, panels B and D), where the knee is often used as the optimal regularization parameter. This ensures that the least-squares solution is neither dominated by contributions from data errors (too little regularization - high solution norm) nor errors due to poor solution fit to the data (too heavily regularized - high residual norm). We desire that regularization creates a better-conditioned problem and mitigates the effects of random noise in data, thereby providing a result that approximates the true solution. Stated differently, regularization in ill-conditioned inverse problems helps in diminishing contributions from small singular values of the inverse matrix ($\bf{F^T F}$, in equation~(\ref{RLSinverse})) that otherwise tend to either result in large-amplitude solutions or leave inferences excessively sensitive to noise. The goal is then stated - for each flow coefficient, minimize the sum of squared differences between data and model, called the residual norm, while penalizing a high solution norm.  
Owing to its past success \citep[][]{Antia1994}, we parametrize velocity $w_{st}^{\sigma}$ in a cubic $B$-spline basis (which we denote using $X$ to avoid confusion with $B$-coefficients) with $70$ knots,
\begin{equation}\label{bspline}
    w_{st}^{\sigma}(r) = \sum_{k}\beta^{k\sigma}_{st}X_{k}(r).
\end{equation}
For every triplet ($s,t,\sigma$), the least-squares problem (without regularization) is posed as (using the above, Equation~(\ref{bspline}))
\begin{equation}\label{leastsquares}
\begin{gathered}
\int_{\odot} dr\:\Theta_{st}^{st}(r;n,\ell)\;w_{st}^{\sigma}(r) =  B_{st}^{\sigma}(n,\ell),\\
\sum_{k}\Bigg[\int_{\odot} dr\:\Theta_{st}^{st}(r;n,\ell)\;X_{k}(r)\Bigg]\;\beta^{k\sigma}_{st}=  B_{st}^{\sigma}(n,\ell).
\end{gathered}
\end{equation}
Here we neglect the $\sum\limits_{s',t'}$ and only retain the self-leakage term $(s',t')=(s,t)$ for simplification. Our motivation for this simplification stems from encouraging results in our prior work - \citet{Mani2020} (Appendix C, and Figure 2, panels C and D) tested this simplified inversion of synthetic $B$-coefficients, without factoring in leakage from neighboring flow modes in the least-squares expression. The final inference of velocities matched well with the input flow profile.\\
Denoting the integral on the left-hand side of equation~(\ref{leastsquares}) by $\boldsymbol{F}$, Equation~(\ref{leastsquares}) is rewritten in matrix form, 
\begin{equation}\label{RLSinverse}
\begin{gathered}
    \boldsymbol{F}\cdot\boldsymbol{\beta} = \boldsymbol{B}, \\
    \boldsymbol{\Big(F^T F\Big)} \cdot\boldsymbol{\beta} = \boldsymbol{F^T B}, \\
    \boldsymbol{\beta} = \boldsymbol{\Big(F^T F\Big)^{-1}}\cdot\boldsymbol{F^T B}.\\
\end{gathered}
\end{equation}
We redefine $\boldsymbol\beta$ to include regularization, i.e.,
\begin{equation}
    \boldsymbol{\beta} = \boldsymbol{Q}\cdot\boldsymbol{B},
\end{equation}
where $\boldsymbol{Q} = \Big( \boldsymbol{F^T F} + \lambda \boldsymbol{I} \Big)^{-1}\boldsymbol{F^T}$, $\lambda$ is the regularization parameter,  and $\boldsymbol{I}$ the Identity matrix. RLS describes a family of methods to solve least-squares problems. The choice of method is as important as how the choice of regularization and the associated parameter. For instance, one can construct a different smoothing matrix (instead of $\bf{I}$) that penalizes higher derivatives in addition to an optimized norm. Here, we use the popular Tikhonov regularization, according to which our minimization problem is given by $\sum\limits_{k}\;||F\beta-B||_2 + \lambda||\beta||_2$, where $||\cdot||_2$ stands for $L_2$ norm. The proper choice of $\lambda$ is, as described earlier, given by the knee of the L-curve, as shown in Figure~\ref{fig_avgk}B. \\
For the depth $r_0$, this allows us to rewrite Equation~(\ref{bspline}) as
\begin{equation}\label{wst_bcoeff}
\begin{gathered}
    w_{st}^{\sigma}(r_0) = \sum_{k}\;\Bigg[\sum_{n\ell}Q^{k}_{n\ell}B_{st}^{\sigma}(n,\ell)\Bigg]\;X_{k}(r_0), \\
    = \sum_{n\ell}\alpha_{n\ell}(r_0)B_{st}^{\sigma}(n,\ell),\\
\end{gathered}
\end{equation}
where $\alpha_{n\ell}(r_0) = \sum_k Q^{k}_{n\ell} X_{k}(r_0)$. These $\alpha$ are the RLS inversion coefficients, permitting us to express velocity as a linear combination of observations and to also construct averaging kernels (see Figure~\ref{fig_avgk}A).
We estimate the noise model and subtract it from the measured signal power; Equation~(\ref{wst_bcoeff}) is thus rewritten as 
\begin{equation}\label{flowfinalrls}
    |w_{st}^{\sigma}(r_o)|^2 \approx  |\sum_{n\ell}\alpha_{n\ell}(r_o)B_{st}^{\sigma}(n,\ell)|^2 - \sum_{n\ell}\alpha_{n\ell}(r_o)^2\epsilon_{st}^{\sigma}(n,\ell). 
\end{equation}

\section{Results and discussion}\label{results}
Convective power is retrieved as the (small) difference between two large numbers: the observed $B$-coefficient power and the  theoretically computed background noise.  
The left-hand side of Equation~(\ref{wst_bcoeff}), $|w_{st}^{\sigma}(r_0)|^2$, is a power-spectrum function of 4 variables $s,t,\sigma$ and $r_0$, and by construction, a positive quantity.

\begin{figure}
\subfloat{\includegraphics[width = 6.2in]{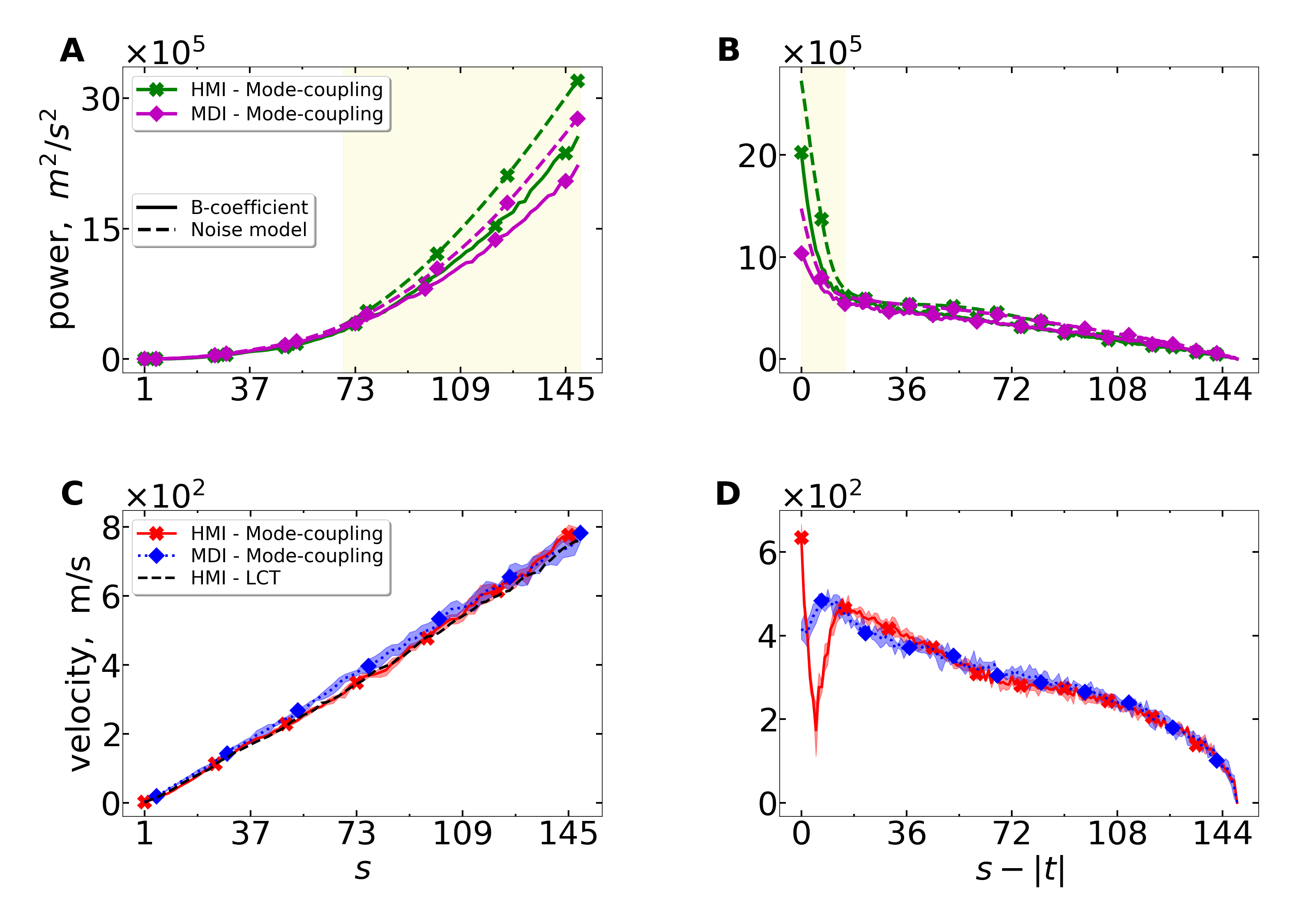}}
\caption{Panels A and B show, for 4 yr HMI (green) and MDI (magenta) data each, inverted using RLS, the B-coefficient power (solid) and the noise power (dashed), in units of $m^2/s^2$ - first and second terms on the RHS of Equation~(\ref{flowfinalrls}) - averaged using Equations~(\ref{eqPs}) and (\ref{eqPdiff}), respectively. The region highlighted in yellow is where noise power exceeds B-coefficient power. Panels C and D show, for 4 yr HMI (red) and MDI (blue) data each, the final velocities, Equation~(\ref{flowfinalbeta}), similarly averaged. The columns and rows share x- and the y-axis labels, respectively. The rows also share the plot legends. Panel C shows that for appropriate regularization, we can obtain an excellent match between mode-coupling inversions and LCT. Note the absence of extra power at supergranular scales ($s\approx120$). LCT is not shown in panel D. The shaded region in the curves of panels C and D denotes $\pm1\sigma$ error.}
\label{fig_ps}
\end{figure}

However, the noise model \citep[see][section 5.1]{Han2018} is incomplete because the leakage matrices do not account for centre-to-limb effects \citep[i.e., the ``shrinking Sun",][]{Duvall2009,Zhao2013}, differential rotation etc., which may contribute to small errors in noise model and therefore translate correspondingly to errors in estimating convective power. We realize its shortcomings in the inversion where we often find the power-spectrum to be negative, i.e., the noise power exceeds measurements: $\sum_{n\ell}\alpha_{n\ell}(r_o)^2\epsilon_{st}^{\sigma}(n,\ell)>|\sum_{n\ell}\alpha_{n\ell}(r_o)B_{st}^{\sigma}(n,\ell)|^2$ see panels A and B, Figure~\ref{fig_ps}, region shaded in yellow) - which occurs in these measurements when $s-|t| \gtrsim 10$. We attempt to overcome this issue by seeking agreement of $\evalat{\sqrt{\sum\limits_{t,\sigma}s(s+1)|w_{st}^{\sigma}(r)|^2}}{r=0.995R_{\odot}}$ with velocities obtained from LCT. That is, we subtract the maximum possible fraction of noise from the measurements such that the power spectrum remains positive and appropriately regularize such that velocity as a function of $s$ from mode-coupling and LCT match. We empirically find the fraction of noise power that can be subtracted to be $0.6$, i.e.,
\begin{equation}\label{flowfinalbeta}
    |w_{st}^{\sigma}(r_o)|^2 \approx  |\sum_{n\ell}\alpha_{n\ell}(r_o)B_{st}^{\sigma}(n,\ell)|^2 - \gamma\sum_{n\ell}\alpha_{n\ell}(r_o)^2\epsilon_{st}^{\sigma}(n,\ell). 
\end{equation}
with $\gamma=0.6$.

Using the definition of power spectra $P_{st}^{\sigma}(r) = s(s+1)|w_{st}^{\sigma}(r)|^2$ in keeping with \citet{Mani2020} and \citet{HanSciA2020}, we average the velocity amplitudes over its dependent variables in two ways as below:
\begin{equation}\label{eqPs}
    P(s,r_0) = \sqrt{\sum\limits_{t,\sigma}P_{st}^{\sigma}(r_o) = \sum\limits_{t,\sigma}s(s+1)|w_{st}^{\sigma}(r_o)|^2},
\end{equation}
\begin{equation}\label{eqPdiff}
    P(s',r_0) = \sqrt{\sum\limits_{\sigma,s,t}P_{st}^{\sigma}(r_o) = \sum\limits_{\sigma,s,t}s(s+1)|w_{st}^{\sigma}(r_o)|^2}.
\end{equation}
where in equation~\ref{eqPdiff}, the summation on $s$ and $t$ are such that $s-|t|=s'$. We plot the velocity distribution (see Figure~\ref{fig_ps}) - $\sqrt{s(s+1)|w_{st}^{\sigma}(r)|^2}$ - at the depth $0.995R_{\odot}$ for these definitions. The observed velocity monotonically decreases with increasing spatial scale (panel C), in contrast to several hydrodynamic simulations and models \citep[e.g.][]{Miesch2008, Lord2014, Hotta2016}. The absence of a peak in power at the supergranular scale ($s\approx120$) leads us to conclude that supergranules are dominantly poloidal \citep[see][]{Langfellner2014,Langfellner2015}. Panel D shows almost a linear drop-off in amplitude for $s-|t| \gtrsim 10$ suggesting comparable amount of power in sectoral and non-sectoral modes.

\begin{figure}
\subfloat{\includegraphics[width = 7in]{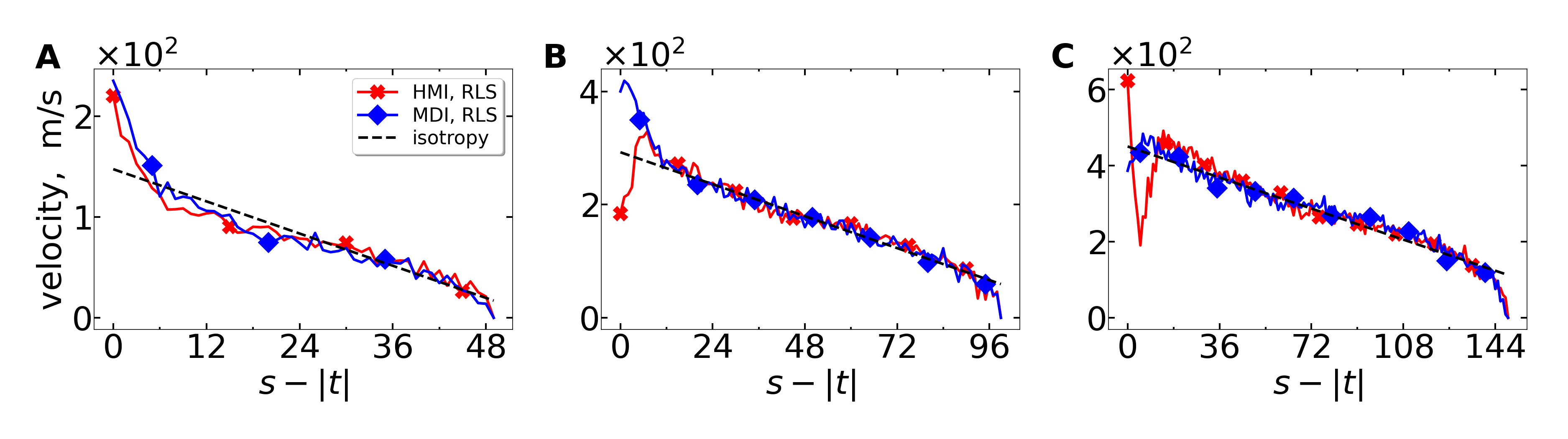}}
\caption{Velocity as a function of $s-|t|$ for 1 yr HMI (red) and MDI (blue), each inverted using RLS. From left to right, the upper limit in $s$ over which averaging is performed - see Equation~(\ref{eqPdiff}) - is $49$, $99$ and $149$, respectively. Indeed for low $s-|t|$, sectoral modes contain excess power. As we extend the spatial scale to include higher wavenumbers, we find that, at the surface, the flow becomes more isotropic, i.e., from left to right, power is more uniformly distributed across all harmonics. Sectoral mode power is greater than that of tesseral or zonal modes, simply because the set of available $(s,t)$ to sum over falls linearly with increasing $s-|t|$ (shown as the nominal dashed line). Note that panel C of this figure is the same as panel D of Figure~\ref{fig_ps}}
\label{fig_isotropy}
\end{figure}

\begin{figure}
\subfloat{\includegraphics[width = 6.2in]{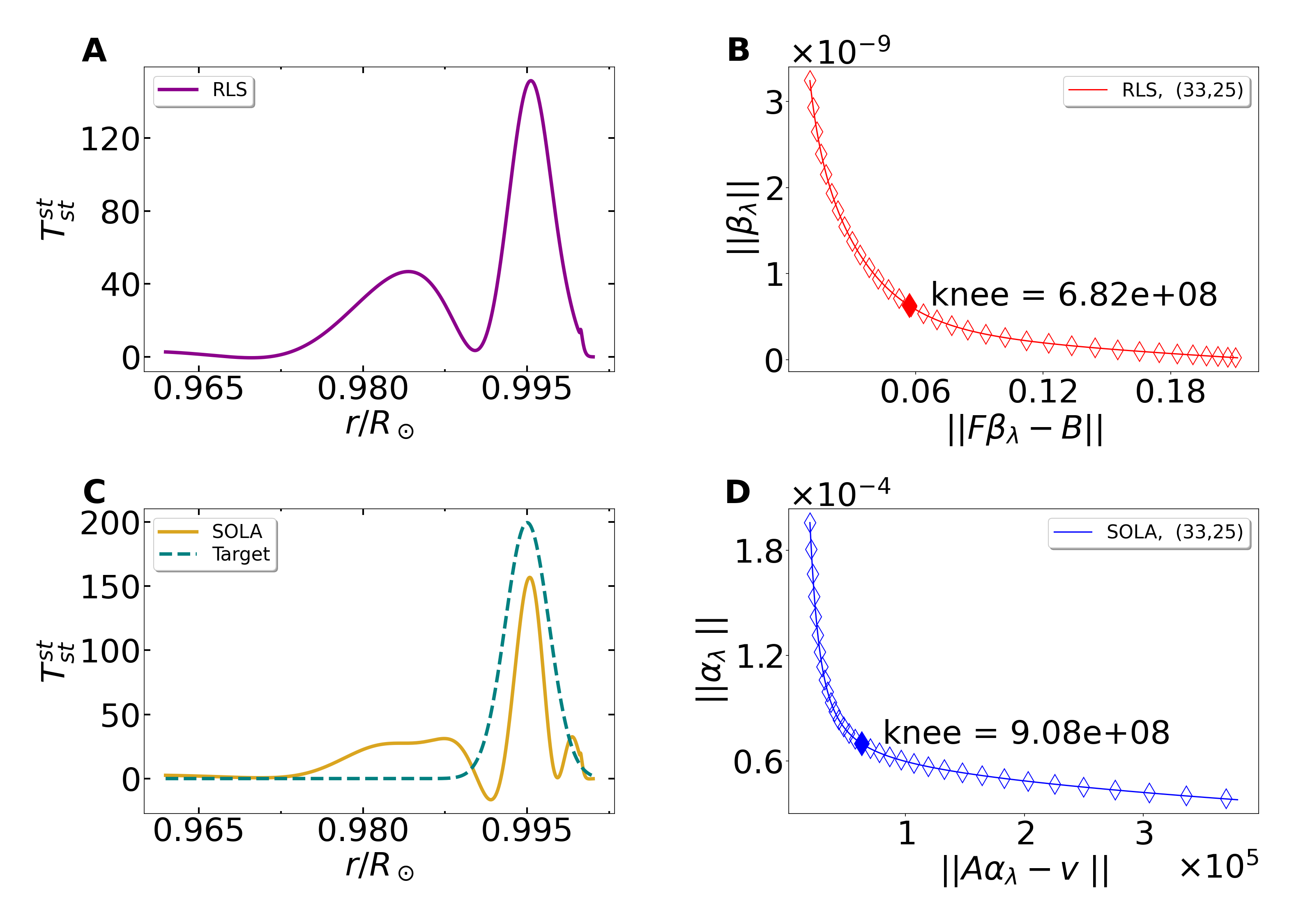}}
\caption{Averaging kernels at $r=0.995R_\odot$ (panels A and C, RLS and SOLA, respectively) and the L-curve along with the knee marked (panels B and D, RLS and SOLA, respectively) - this is the optimal choice for the regularization parameter $\lambda$. Shown here are the curves for the convective mode $(s,t)=(33,25)$. Optimal regularization for a well localized averaging kernel and a minimal residual has to be achieved for each $(s,t)$ in the analysis. In SOLA, the trade-off can be better understood as being between optimally matching the averaging kernel with the target function on one hand and obtaining velocities with appropriate norm on the other.}
\label{fig_avgk}
\end{figure}

Figure~\ref{fig_isotropy} shows the changing geometry of the flow as more and more wavenumbers are included in the averaging formula described in Equation~(\ref{eqPdiff}). Since $|t|\le s$, $s-|t| =k$ has $s_{\rm max}-k$ modes (excluding $t=0$ and $s=0$ modes) and the power correspondingly changes. Thus, in a perfectly isotropic system where each mode reports the exact same power, we would expect the power to follow a linear declining relationship, from its maximum at $s-|t|=0$ to zero at $s-|t|=s$. Keeping this in mind, to characterize the shape of the flow, we define a measure of `isotropy' as the deviation from the expected linear baseline. The more enhanced or depleted the power is in any segment of $s-|t|$ in comparison the nominal linear power variation, the less isotropic the flow. The lack of isotropy implies that flows are more or less vigorous in different parts of the sphere. In panel A, where the sum is over $s\in[1,49]$, power dominates in sectoral modes, i.e., $\rightarrow$ $s-|t|\sim0$ \citep[e.g,][]{HanSciA2020}. This would imply that flows are preferentially vigorous in the equatorial regions. As we average over a larger range of $(s,t)$, as in panels B ($s\in[1,99]$) and C ($s\in[1,149]$), we see the preference for sectoral enhancement vanishing and the power distributed uniformly across all wavenumbers - an indication that the flow is becoming more isotropic and possibly uniformly distributed across the spherical surface. We are however cautious about strongly interpreting the results of Figure~\ref{fig_isotropy}. Panels B and C, where higher wavenumbers are included, exhibit signs of imperfect noise subtraction at low $s=|t|$ (for the HMI data), in line with the yellow shaded regions in Figure~\ref{fig_ps}A and B. Velocity inferred from HMI data (red curve) in panel C at low $s-|t|$ shows deviation from the nominal isotropy baseline that may not be consistent with our interpretation of the flow being isotropic. But the large fluctuations around the isotropy line are possibly from challenges in modelling noise associated with the high wavenumbers. More evidence of problems in theoretically derived mode-coupling noise has recently come to light \citep{Wood2021}.

The Rossby number, which is the ratio of the inverse convective turnover timescale to the rotation rate, quantifies the influence of rotation on the fluid dynamics of the system, i.e., small Rossby numbers correspond to rotationally constrained flow and vice versa.
Supergranules \citep[$U\approx300m/s$, $L\approx30-35$Mm, $Ro\sim O(1)$, see][]{Rincon2018} correspond to rotationally balanced flows; their poloidal flows are observed to show a latitudinal variation \citep[][]{Lisle2004,Nagashima2011,Langfellner2015}, a deviation from isotropy that can be ascribed to Coriolis force, whereas properties of granules \citep[$U\approx3$ km/s, $L\approx1-1.5$Mm, $s\sim2500$; these values are taken from][]{Hathaway2013, Hathaway2015}, with Ro$\sim\mathcal{O}(10^3)$, are largely uniform over the entire solar disk. In fact, it has also been noted that supergranules have a preferred sense of vorticity in each hemisphere \citep[clockwise (anticlockwise) in norther (southern) hemisphere; see][]{Duvall2000}. While in this work we only analyze vortical flows up to $s=150$, we believe they similarly approach a state of increasing isotropy as $s_{max}$ is increased from $50$ to $150$ (see Figure~\ref{fig_isotropy}).

\citet{Han2012} showed, through time-distance helioseismology \citep{Duvall1993}, that velocity amplitudes on large scales ($s\sim60$) are at least an order of magnitude lower than predictions from numerical simulations. The mechanisms governing the observed continuous decrease in power with increasing spatial scales remain as yet unclear. \citet{Lord2014} for instance have suggested that if the convection zone at layers deeper than $0.97R_\odot$  or so were to be adiabatic, the convective spectrum would start decreasing at scales larger than supergranulation, and thereby highlighting supergranules in the spectrum because of power suppression in the adjacent scales. However, supergranular flows show a wave-like dispersion relation \citep{Gizon2003, Hanson2021}, suggesting that it is a privileged scale, not merely arising from a reduced effective depth of the convection zone. \cite{Featherstone2016} observed a peak power scale highlighted for different Rossby numbers in convective simulations, suggesting that deep convection of larger-scale modes are rotationally inhibited, and scales smaller than this enhanced peak scale behave like non-rotating convection, i.e., they are only weakly influenced by Coriolis force. Our results lends credence to the analysis of \citet{Featherstone2016}: Plots shown in Panel C, Figure~\ref{fig_ps}, and in Panels A through C in Figure~\ref{fig_isotropy}, indicate that large-scale power on the surface is low because they are possibly rotationally inhibited, and smaller scales increasingly behave like non-rotating convection, exhibiting increasing isotropy.

Although LCT is used for comparison, we emphasize that it is fundamentally different from seismology. While LCT is successful in measuring surface horizontal velocity fields insofar as the length and time scales of the inferred flow are well separated from those of granules \citep[][]{Rieutord2001}, inversions through seismology makes full use of sensitivity kernels - a product of mode-eigenfunctions - that peak at a few Mm under the surface, permitting inferences of subphotospheric velocities. Additionally, we neglect observations for LCT for latitudes higher than $60^\circ$ in both the hemispheres because of systematics. Since LCT tracks the movement of granules, it implies a specific depth averaging of the large-scale flows that are inferred. This is evidently different from that of seismology. 

\section{Summary}\label{conclusion}
Our analysis points to the idea that, as higher wavenumbers are gathered into the analysis, toroidal flow becomes increasingly isotropic, i.e., power is more uniformly spread across the full range of spherical harmonics. The present work is a continuation of the analysis carried out in \citet{HanSciA2020} and \citet{Mani2020}. The synthetic tests undertaken in the latter included spatial leakage in the kernels computed for toroidal flow, and the result that inversions are independent of the temporal frequency at which kernels are computed served well in saving computing time for the present analysis. As has been noted in \citet{Han2018}, correlated realization noise brought about by leakage is non-trivial to model and there remains work to be done to better understand the missing ingredients in the model and improve estimates of the leakage matrices.

We reported a finding in this document - toroidal flow appears to contain uniform power when smaller spatial scales are analyzed (see Figure~\ref{fig_isotropy}). Figure~\ref{fig_ps}C also confirms part of the result from \citet{Rincon2017} and \citet{Hathaway2015}, that flows at supergranular scale ($s\approx120$) are, at best, weakly toroidal. Future work will focus on appreciating poloidal (horizontal-divergent) flow using mode-coupling and comparing with the velocities obtained here.

P.M. is grateful to Samarth G. Kashyap for several clarifying discussions regarding data analysis. Sashi Kiran Mahapatra was immensely helpful in resolving issues relating to computation. P.M and S.M.H. acknowledge that LCT data was generated by Bjoern L\"{o}eptien and provided by the authors of \citet{Chris2020}. The knee of the L-curve was found using the python module given in the github repository \url{ https://github.com/arvkevi/kneed }- Finding a “Kneedle” in a Haystack: Detecting Knee Points in System Behavior. 

\appendix
\section{SOLA Inversions}\label{SOLA Inversions}
As an illustration, we only show 1 yr HMI inversion using SOLA at $r=0.995R_\odot$.
In SOLA, the goal is to find a set of coefficients $\alpha_{n\ell}(r_0)$ for the depth $r_0$ such that the weighted sum of the kernels $\alpha_{n\ell}(r_0)\Theta_{st}^{st}(r;n,\ell)$ is left sensitive only to the flow around that depth. The `averaging' kernel $T$ is thus given by
\begin{equation}\label{avgkleak}
    T_{st}^{st}(r,r_o) = \sum_{n\ell}\alpha_{n\ell}(r_o)\Theta_{st}^{st}(r;n,\ell).
\end{equation} 
The flow velocity for a noiseless measurement at $r_0$ is given by
\begin{equation}\label{avgkdelta}
    w_{st}^{\sigma}(r_o) \approx  T_{st}^{st}(r,r_o) w_{st}^{\sigma}(r).
\end{equation}

\begin{figure}
\subfloat{\includegraphics[width = 6.2in]{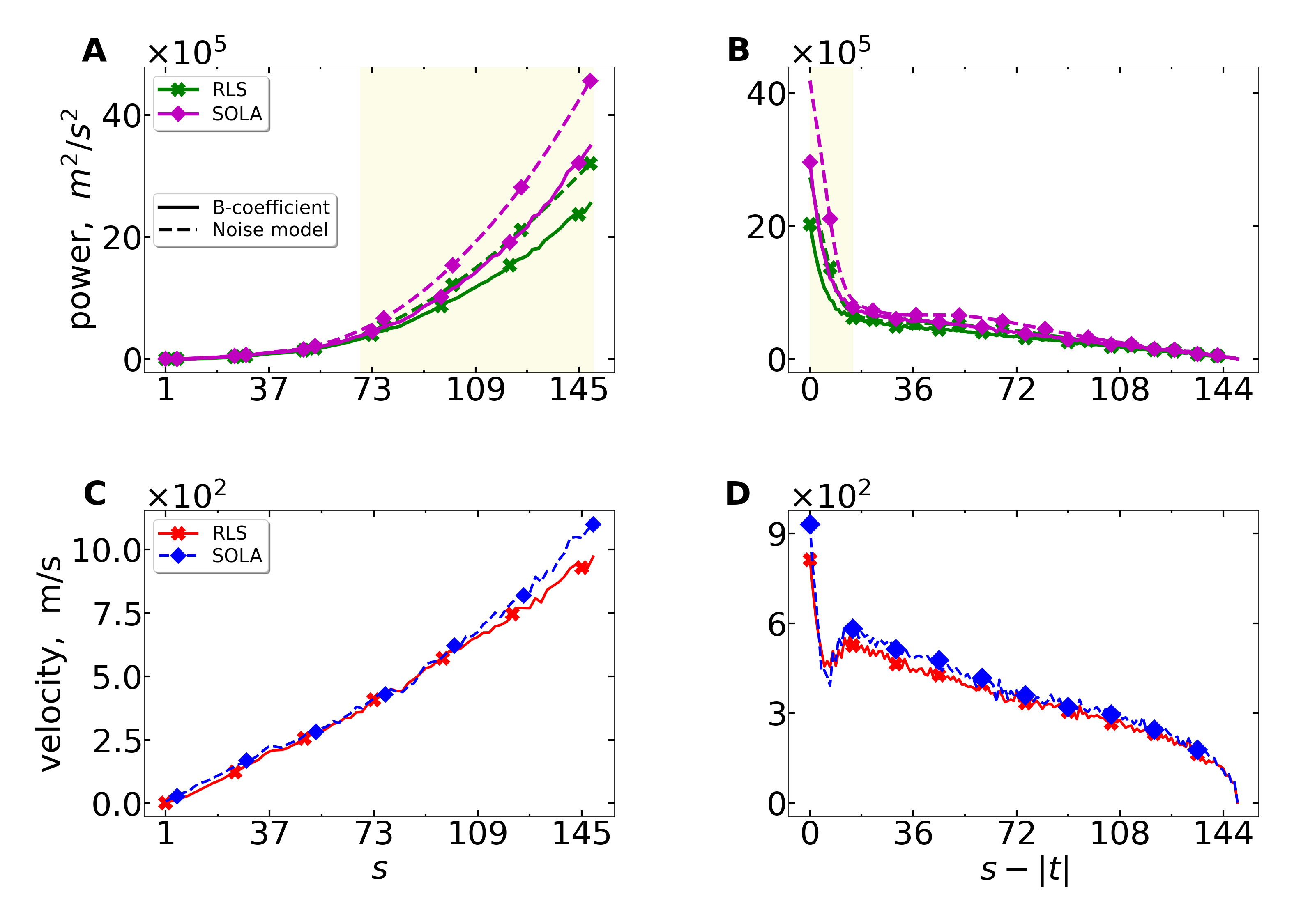}}
\caption{Panels A and B show, for 1 yr HMI data inverted using RLS and SOLA, the B-coefficient power (solid) and the noise power (dashed), in units of $m^2/s^2$ - first and second term in the RHS of Equation~(\ref{flowfinalrls}) - averaged using Equation~(\ref{eqPs}) and (\ref{eqPdiff}), respectively. The region highlighted in yellow is where the noise propagation becomes problematic - noise power exceeds B-coefficient power. Panels C and D show the final velocities, Equation~(\ref{flowfinalbeta}), similarly averaged. The figure columns and rows share the x- and the y-axis labels, respectively. The figure rows also share the plot legends. Here we set $\gamma=0.5$ for comparison. }
\label{fig_sola_comp}
\end{figure}

Absent noise, the closer the averaging kernel is to a $\delta$ function, the more accurate the inference of the flow at that depth. Presence of noise brings about a trade-off between minimizing noise contribution and spatially localizing the averaging kernel \citep{PandT1994}. Hence, the optimization problem in the inversion is stated such that the averaging kernel resembles a normalized `target' Gaussian $\mathcal{T}(r,r_0)$ with centre at $r_0$ and a suitable width (see Figure~\ref{fig_avgk}C):
\begin{equation}
    \chi_{st} = \frac{1}{2} \int_{\odot} \;dr\;\Big[ \mathcal{T}(r,r_0)-T_{st}^{st}(r,r_o)\Big]^2. 
\end{equation}
This leads to a matrix equation
\begin{equation}
\begin{gathered}
\boldsymbol{A}\cdot\boldsymbol{\alpha} = \boldsymbol{v}, \\
\boldsymbol{\alpha} \approx (\boldsymbol{A} + \lambda\boldsymbol{I})^{-1}\cdot\boldsymbol{v},
\end{gathered}
\end{equation}
where $A = \int_\odot dr\;\Theta_{st}^{st}(r;n,\ell)\Theta_{st}^{st}(r;n',\ell')$ and $v = \int_\odot \Theta_{st}^{st}(r;n,\ell)\mathcal{T}(r,r_0)$. Regularization finds an appropriate balance (see Figure~\ref{fig_avgk}D) for the minimization of $\sum\limits_{n\ell}\;||A\alpha-v||_2 + \lambda||\alpha||_2$, where $||\cdot||_2$ stands for $L2$ norm. 
Therefore, using Equation~(\ref{Bst}) and Equation~(\ref{avgkleak}), Equation~(\ref{avgkdelta}) is restated as 
\begin{equation}\label{flowbcoef}
    w_{st}^{\sigma}(r_o) \approx  \sum_{n\ell}\alpha_{n\ell}(r_o)B_{st}^{\sigma}(n,\ell).
\end{equation}
We estimate the noise model and subtract that from the measured signal power; equation~(\ref{flowbcoef}) is thus rewritten as 
\begin{equation}\label{flowfinalsola}
    |w_{st}^{\sigma}(r_o)|^2 \approx  |\sum_{n\ell}\alpha_{n\ell}(r_o)B_{st}^{\sigma}(n,\ell)|^2 - \sum_{n\ell}\alpha_{n\ell}(r_o)^2\epsilon_{st}^{\sigma}(n,\ell). 
\end{equation} 
We find the maximum fraction noise that can be subtracted using SOLA to be $0.5$. We thus show comparison of SOLA and RLS, where for both, we have subtracted $0.5$ noise - see panels C and D, Figure~\ref{fig_sola_comp}.

\end{document}